\documentclass[namedreferences]{solarphysics}

\usepackage[hyperref,optionalrh]{spr-sola-addons}
\usepackage{graphicx}        
\usepackage{color}           
\usepackage{breakurl}        

\chardef\us=`\_

\begin{document}

\begin{article}
\begin{opening}

\title{On the synchronizability of Tayler-Spruit and 
Babcock-Leighton type dynamos}

\author[corref,email={F.Stefani@hzdr.de}]{\inits{F.}\fnm{F.}~\lnm{Stefani}}
\author{\inits{A.}\fnm{A.}~\lnm{Giesecke}}
\author{\inits{N.}\fnm{N.}~\lnm{Weber}}
\author{\inits{T.}\fnm{T.}~\lnm{Weier}}
\address{Helmholtz-Zentrum Dresden -- Rossendorf, P.O. Box 510119,
D-01314 Dresden, Germany}

\runningauthor{F. Stefani {\it et al.}}
\runningtitle{On the synchronizability of Tayler-Spruit and 
Babcock-Leighton type dynamos}

\begin{abstract}The solar cycle appears to be remarkably
synchronized with the gravitational torques exerted by
the tidally dominant planets Venus, Earth and Jupiter.
Recently, a possible 
synchronization mechanism was proposed that
relies on the intrinsic helicity oscillation
of the current-driven Tayler instability which
can be stoked by tidal-like perturbations
with a period of 11.07 years.
Inserted into a simple $\alpha-\Omega$ dynamo model
these resonantly excited helicity oscillations
led to a 
22.14 years dynamo cycle. Here, we assess
various alternative mechanisms of synchronization. 
Specifically we study a simple time-delay model of
Babcock-Leighton type dynamos and ask whether 
periodic changes of either the minimal amplitude 
for rising toroidal flux tubes or the $\Omega$ effect 
could eventually lead to synchronization.
In contrast to the easy and robust synchronizability of
Tayler-Spruit dynamo models, our  answer for those 
Babcock-Leighton type models is less propitious.

\end{abstract}
\end{opening}
\section{Introduction}

Despite its long history, which traces back to
\cite{Wolf1859},
the idea of planetary influence on the solar dynamo 
is widely considered  as marginal, if not
''astrological''. There are indeed good reasons for 
skepticism: the gravitational forces 
exerted by the planets are tiny when compared 
to the intrinsic buoyancy and Coriolis  
forces that are believed to govern the 
solar dynamo \citep{Callebaut2012}.
The tidal accelerations of a few 
10$^{-10}$\,m/s$^2$,  which produce
a tidal height of the order of 1\,mm at the tachocline
\citep{CondonSchmidt1975},
seem - at first glance - ridiculous when asking for
possible influences on the solar dynamo.

Yet, there are remarkable correlations of the 
solar cycle with planetary orbits. 
This applies, in particular, to the apparent synchronization
\citep{Bollinger1952,Takahashi1968,Wood1972,Opik1972,CondonSchmidt1975,Grandpierre1996,Hung2007,Wilson2013,Okhlopkov2014,Okhlopkov2016}
of the solar cycle with the 11.07 years conjunction 
cycle of Venus, Earth and Jupiter, which are the
three dominant tide producing planets
\footnote{The ''generous omission'' of Mercury, 
whose tidal effect is nearly the same as that of 
Earth, but whose 88 days  revolution 
period is often considered as ``so short
that its influence appears only as an 
average, non-fluctuating factor...'' \cite{Opik1972}, 
might be another argument for skeptics. 
However, it could also be worthwhile
to re-analyze the 50-80 years sub-band of the 
Gleissberg cycle as identified by \cite{Ogurtsov2002}
in the light of the 66.4 years period
of the four-fold co-alignment of Mercury, Venus, Earth
and Jupiter \citep{Verma1986}.}. 
The average solar cycle duration of 
$(2008.9-1610.8)/36=11.06$ years, derived from the data 
of the last 36 cycles  \citep{Richards2009,Li2011}, 
indicates an astonishing coincidence. Even more 
remarkable is the recent finding of 
\cite{Okhlopkov2016} that the 
synchronization may have lasted for the 
last 50 cycles.
Further to this, fossil records suggest that 
the solar cycle has been amazingly stable for at 
least the last 290 million 
years: the cycle length during the 
early Permian, e.g., was recently estimated as 
10.62 years \citep{Luthardt2017}.
Hence, one might reconsider the tidal height 
$h_{\rm tidal} \approx 1$\,mm and
ask whether this is indeed as irrelevant as it looks like.
Given the huge gravitational acceleration at the 
tachocline of $g \approx 500$\,m/s$^2$ \citep{Wood2010}, we find  
an equivalent velocity of 
$v \sim (2 g h_{\rm tidal})^{1/2} \approx 1$\,m/s which is 
not at all negligible, as 
already  noted by \cite{Opik1972}. 

While such a ``hard'' {\it synchronization} of the
basic Hale cycle with planetary tidal 
forces was advocated by only a few researchers, much 
more interest was dedicated to various kinds of ``soft'' 
planetary {\it modulation} of that cycle (whose length 
is usually believed to be determined by 
intrinsic solar parameters \citep{Charbonneau2010,Cameron2017}). 
Intriguing connections have been
found between various periodicities of the
solar magnetic field (Suess-de Vries, Hallstadt, Eddy etc.)
and corresponding
planetary constellations
\citep{Jose1965,Charvatova1997,Abreu2012,Wolf2010,Scafetta2010,Scafetta2014,McCracken2014,Cionco2015,Scafetta2016}.
As an example, \cite{Abreu2012} had 
revealed synchronized cycles in proxies of the solar activity 
and the planetary torques, with periodicities that 
remain phase-locked over 9400 years. While
still under scrutiny \citep{Cameron2013,Poluianov2014,Abreu2014},
any such relationship - if confirmed - would have 
important consequences for the predictability not only 
of the solar dynamo but, very likely, 
of the terrestrial climate, too
\citep{Hoyt1997,Gray2010,Solanki2013,Scafetta2013,Ruzmaikin2015,Soon2016}.

Returning to the problem of ``hard'' synchronization,
we have recently tested a physical
mechanism that seems promising for explaining it 
\citep{Stefani2016}. 
We set out from a rarely
discussed type of stellar dynamo models, in which
the poloidal-to-toroidal field transformation is
traditionally ensured by the $\Omega$ effect, 
while the
toroidal-to-poloidal transformation
starts only when the toroidal field itself becomes 
unstable to a non-axisymmetric 
current-driven instability.
Early versions of such a  
dynamo mechanism were discussed
by \cite{Ferrizmas1994} and \cite{Zhang2003}, and
auspiciously applied to explain grand minima 
in terms of on-off intermittency \citep{Schmitt1996}.
The underlying kink-type Tayler instability (TI) 
had been theoretically treated by many authors 
\citep{Tayler1973,Pittstayler1985,Gellert2011,Ruediger2013,Ruediger2015,Stefani2015},
and was recently also 
observed in a liquid metal experiment 
\citep{Seilmayer2012}.
Based on this TI, a version of a non-linear 
dynamo mechanism had been proposed \citep{Spruit2002} 
which is now known as the ``Tayler--Spruit dynamo''.
This first version was soon criticized 
by \cite{Zahn2007}
who argued that the non-axisymmetric ($m=1$) TI mode 
would produce the ``wrong''  poloidal field, 
being unsuitable for regenerating the dominant 
axisymmetric ($m=0$) toroidal field. Fortunately, the same 
authors offered a possible rescue 
for the Tayler-Spruit dynamo concept provided 
that the $m=1$ TI would produce an 
$\alpha$ effect with some $m=0$ component. 

The emergence 
of such a TI-related $\alpha$ effect is far from trivial, though.
For comparable large values
of the magnetic Prandtl number [$Pm$], 
\cite{Chatterjee2011,Gellert2011,Bonanno2012,Bonanno2017} 
found evidence for spontaneous symmetry breaking between 
left- and right-handed TI modes, leading
indeed to a finite value of $\alpha$.
Things are different, however,  
for low $Pm$ (as typical for the solar tachocline)
for which we observed 
a tendency of the TI to produce 
oscillations of the helicity and the $\alpha$-effect
related to it \citep{Weber2013,Weber2015}.
The first result of \cite{Stefani2016} was 
that those oscillations between  
left- and right-handed $m=1$ TI modes
are very susceptible to  $m=2$ perturbations, which 
could explain their easy synchronizability with
tidal forces as exerted by planets.

Appropriately parametrized, this resonant behaviour
of $\alpha$ was then implemented into a 
simple zero-dimensional 
$\alpha-\Omega$ dynamo model which turned out 
to undergo oscillations with period doubling.
In summary, we found that an 11.07 year 
tidal-like oscillation may lead to a resonant 
excitation of a 11.07-year oscillation of the 
TI-related $\alpha$-effect, and thereby to a 
22.14 year Hale cycle of the entire dynamo.

We note in passing that the 
notion ``Tayler-Spruit dynamo'', as 
used above,
is not exactly correct for our modified 
model which comprises, 
besides of the resonantly excited oscillatory 
part of $\alpha$, also some small, but non-zero constant part
(subjected only to some standard type of $\alpha$-quenching).
Interestingly, the product of this 
constant part of $\alpha$ with $\Omega$ had to be
positive to make the dynamo working, 
while a negative product led to decaying 
solutions.
In a slightly extended model \citep{Stefani2017} 
we further showed that this positive product
can even provide the correct equator-ward 
direction of the 
butterfly diagram of sun-spots, in pleasant contrast 
to what one would
naively expect from the Parker-Yoshimura sign rule  
\citep{Parker1955,Yoshimura1975,Pipin2013}.

In the light of such promising features of a tidally 
synchronized solar dynamo model of  Tayler-Spruit type
(with the semantic caveat noticed above), 
in this paper we ask for alternative synchronization 
mechanisms which are closer to the more widely accepted 
concept of flux-transport dynamos \citep{Charbonneau2010}. In those models, the 
Babcock-Leighton mechanism 
\citep{Babcock1961,Leighton1964} 
interprets the generation of poloidal field by 
the stronger diffusive cancellation 
of the (closer to the equator) leading sunspots compared with 
that of the trailing (farther from the  equator) spots. 
This leads to 
a spatially separated, or flux-transport 
type of dynamo, which is 
also known to exhibit correct 
butterflies if combined with an 
appropriate meridional circulation \citep{Choudhuri1995}.

Specifically, we will investigate two 
models of putative planetary influences
on flux-transport dynamos.
The first model relies on periodic tidal perturbations
of the adiabaticity in the  tachocline region, which  is 
crucial for the
storage capacity of magnetic flux tubes before they
are set free to erupt \citep{Abreu2012}. This perturbation
will be emulated as a periodic change of the 
minimum magnetic field beyond which magnetic flux tubes 
are allowed to rise.

The second model traces back to an idea of 
\cite{Zaqarashvili1997} who had explained the 22 years cycle 
as an Alfv\'en wave excited 
via parametric 
resonance  from a 11 years period change
of differential rotation, which - in his view -
could rely on the motion of the sun around the
Sun-Jupiter common mass center. 
For this model
to work it had to pre-suppose a significant 
poloidal field of fossil origin.
An 11 years oscillation of the differential rotation 
is indeed known from measurements \citep{Brown1989,Howe2009},
although it is usually explained in terms 
of back-reaction of the dynamo field on the flow, 
either by the Malkus-Proctor effect \citep{Malkus1975}
or by so-called $\Lambda$-quenching \citep{Kitchatinov1994}.

As in \cite{Stefani2016}, we will refrain from any
expensive higher dimensional simulations and restrict 
ourselves to simple, zero-dimensional dynamo models. 
As a basis we will utilize the time-delay concept 
introduced by \cite{Wilmotsmith2006} 
which appears to be particularly 
suited for our purposes. It implements two time delays 
into the dynamo cycle, one representing the rise time
of flux-tubes which transports
toroidal field from the tachocline to the 
working site of the $\alpha$-effect, the other one 
representing the time needed for the meridional
circulation to transport poloidal field  
back to the working site of the $\Omega$-effect.

Actually, both synchronization mechanisms will be studied for three 
paradigmatic regimes distinguished  by 
the ordering of the involved time scales, namely a 
diffusion-dominated, a flux-transport dominated, and an 
intermediate regime.
In either case, we will further distinguish between negative
and positive products of $\alpha$ and $\Omega$ which 
typically lead to oscillatory or 
pulsating behaviour, respectively. In higher-dimensional 
models this sign is known to determine 
the direction of the butterfly 
diagram \citep{Parker1955,Yoshimura1975,Pipin2013}: 
this issue will not be discussed here.

However, before entering these new topics, we will recall
the synchronization mechanism based on the 
Tayler-Spruit-type 
dynamo model and discuss, as a leftover question from 
\cite{Stefani2016}, its behaviour with respect to 
variations of the diffusion time.

\section{Synchronization of Tayler-Spruit type dynamos}

In this section we will analyze a further aspect of 
the reduced zero-dimensional
$\alpha$--$\Omega$ dynamo model of \cite{Stefani2016}, consisting
of two coupled ordinary differential equations for
the toroidal and the poloidal field components,
\begin{eqnarray} 
  \frac{{\rm d} b(t)}{{\rm d} t} &=& \Omega a(t) - \frac{b(t)}{\tau} \\
    \frac{{\rm d} a(t) }{{\rm d} t} &=& \alpha(t) b(t) - \frac{a(t)}{\tau} , 
    \label{system_tayler}
   \end{eqnarray}
wherein $a$ represents the poloidal field (specifically, its 
vector potential), and $b$ the toroidal field. 
While in \cite{Stefani2016} the diffusion time had been 
fixed to $\tau=1$ year, it will now be considered as  
variable (for the sake of shortness, we will 
skip the time unit ``year'' in all following 
numerical analyses).

\begin{figure}[!ht]
\includegraphics[width=120mm]{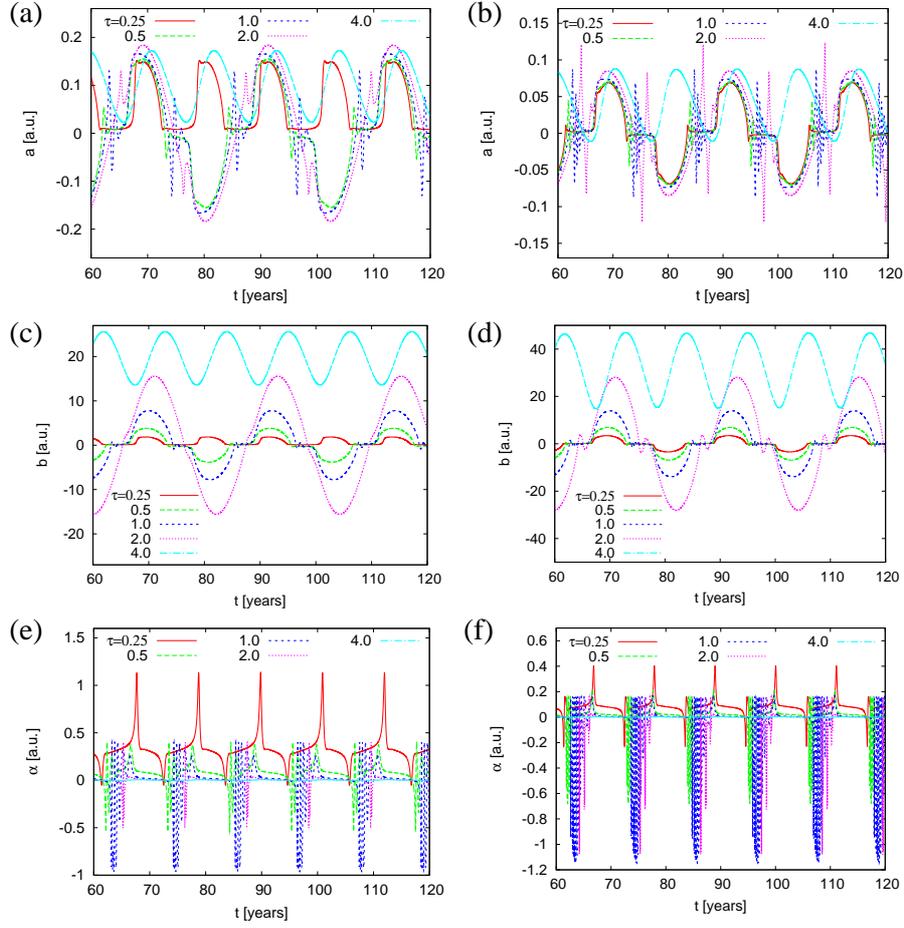}
\caption{Time dependence of $a$ (a,b), $b$ (c,d) and $\alpha$ (e,f)
for the two parameter sets $\Omega=50$, $c=0.4$, $p=8$, and $h=10$ (a,c,e) 
and $\Omega=200$, $c=0.16$, $p=8$, and $h=10$ (b,d,f), and
varying values of $\tau$. For the case (a,c,e), pulsations with an 11.07  
period occur for
a low $\tau=0.25$ and a large $\tau=4$, while for intermediate $\tau$ we
find oscillations with a period of 22.14. For the case (b,d,f), 
pulsations are found only for the large $\tau=4$, while for other
 $\tau$ oscillations occur.
}
\label{Fig:alpha_osci}
\end{figure}

In contrast to the constant value of $\Omega$, which 
represents the induction effect of the differential rotation,
$\alpha$ is supposed to depend on the 
instantaneous toroidal magnetic field:
\begin{eqnarray}   
    \alpha(t) &=& \frac{c}{1+g b^2(t)}
                          +\frac{p b^2(t)}{1+ h b^4(t)} \sin{(2 \pi t/T_{\rm tidal})} \; .
	\label{alpha_tayler}		  
   \end{eqnarray}
Equation (\ref{alpha_tayler}) is motivated as follows: the first term, 
scaled by $c$,
reflects some {\it constant} part that is only quenched, in 
the usual way, by the magnetic-field energy ($b^2$) in 
the denominator. 
The second contribution, scaled by a parameter 
$p$, is {\it periodic} in time and   
emulates the resonant excitation of the $\alpha$-oscillation 
by fixing its period to $T_{\rm tidal}$, but 
maximizing its amplitude at some particular value of 
$|b|$ where the tidal excitation happens to be 
in resonance with 
the intrinsic helicity oscillation of the TI.

As shown previously  \citep{Stefani2016}, this dynamo 
fails to work when either $c$ is set to zero, or the
product of $c$ and $\Omega$ is negative.
Working dynamos come in two guises: 
their field might 
either pulsate (keeping one single sign) with a period 
equal to $T_{\rm tidal}$, 
or oscillate with a period of $2 T_{\rm tidal}$, which 
is the ``desired'' behaviour that was indeed 
found for a wide range of
parameters.

What was left over from \cite{Stefani2016} was an 
assessment of the role of the diffusion time $\tau$ which 
had been
set to 1 for the assumed tidal forcing period of 
11.07. Figure \ref{Fig:alpha_osci} illustrates 
now the results for varying $\tau$. For the parameter choice
$\Omega=50$, $c=0.4$, $p=8$, and $h=10$ (a,c,e), 
pulsations with a 11.07 period occur for 
a low value $\tau=0.25$ as well as for a large value 
$\tau=4$, while for intermediate $\tau$ we
find oscillations with a period of 22.14. 
For the case $\Omega=200$, $c=0.16$, $p=8$, and $h=10$ (b,d,f),
pulsations are found for the largest considered value $\tau=4$, 
while for other $\tau$ oscillations are observed.

\begin{figure}[!ht]
\includegraphics[width=120mm]{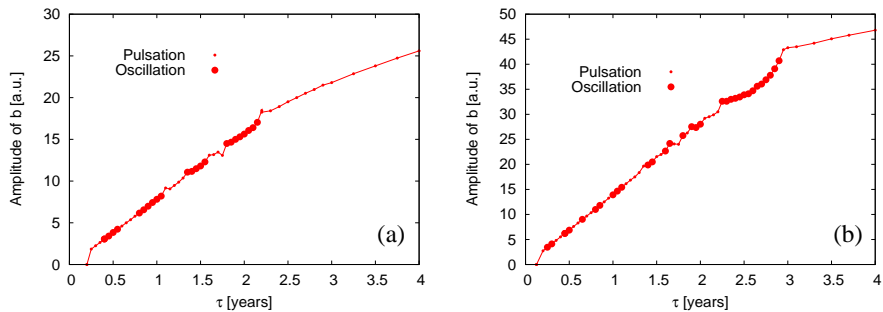}
\caption{Transitions between oscillations and pulsations
for varying the diffusion time $\tau$ for $\Omega=50$, $c=0.4$, 
$p=8$ (a), and $\Omega=200$, $c=0.16$, $p=8$ (b).
}
\label{Fig:behaviour}
\end{figure}

A more detailed analysis reveals, however, a more
complicated behaviour, as 
documented in Figure \ref{Fig:behaviour}. We observe
a sequence of transitions between oscillatory and
pulsatory behaviour, partly with quite narrow bands,
in particular for the higher value $\Omega=200$ (b).

While the oscillatory behaviour is what we are 
usually searching for, one might ask whether the 
pulsations are indeed as unphysical as they 
look like. It is tempting here to 
think about Maunder type grand minima, for which
measurements of $^{10}$Be had indicated a 
rather unperturbed 11 years cycle \citep{Beer1998}, 
possibly connected, however,  with a changed field parity
\citep{Sokoloff1994,Weiss2016,Moss2017}. 
While our zero-dimensional
model cannot decide about the parity 
of 3D-dynamo fields, it might be interesting to 
learn
into what kind of behaviour 
our pulsations  would translate once 
higher dimensional models were applied.
Let us bravely assume for the moment that our pulsation regime
would indeed correspond to the regime of grand minima. We might 
then ask for the details of transitions into, and from, such 
a minimum.

\begin{figure}[!ht]
\begin{center}
\includegraphics[width=80mm]{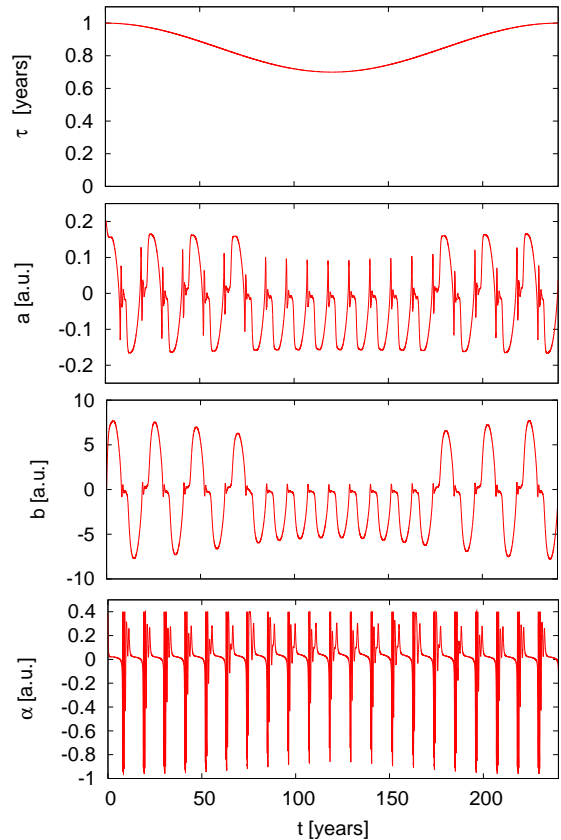}
\caption{Transitions between oscillations and pulsations, and
back, when changing the value of $\tau$. Note the phase
coherence throughout the ``grand minimum''.}
\label{Fig:maunder}
\end{center}
\end{figure}

For $\Omega=50$, $c=0.4$, $p=8$, and $h=10$,
Figure \ref{Fig:maunder} illustrates such a process,
by forcing the diffusion time $\tau$ 
to change between 1  and 0.7, and back.
As can be expected from Figure \ref{Fig:behaviour}a,
the initial oscillations is replaced, at $t\approx 80$,
by a pulsation, but recovers again at $t\approx 170$. 
It is most remarkable that the periodicity 
remains ''on beat'' during these two transitions.

\section{Synchronization of Babcock-Leighton type dynamos}

We turn now to the question whether a similar synchronization
effect, as discussed in the previous section for Tayler-Spruit 
type dynamos, could also
result from appropriate periodic changes in a flux-transport (or 
Babcock-Leighton) dynamo model. After describing the mathematical
model, we will present simulations for three regimes which differ in 
the ordering of the relevant time-scales.

\subsection{The model}

We modify the model of equation system (1,2,3) 
according to 
\cite{Wilmotsmith2006}, who had 
introduced two specific
time delays in their dynamo model. Consider the system
\begin{eqnarray} 
  \frac{{\rm d} b(t)}{{\rm d} t} &=& \Omega(t) a(t-\tau_0) - \frac{b(t)}{\tau} \\
    \frac{{\rm d} a(t) }{{\rm d} t} &=& \alpha(t-\tau_1) b(t-\tau_1) - \frac{a(t)}{\tau}
    \label{system_babcock}
   \end{eqnarray}
with
\begin{eqnarray}   
    \alpha(t) &=& \alpha_0 \frac{1}{4} [1+{\textrm {erf}}( b^2(t)-b^2_{\rm min}(t)) ] 
    [1-{\textrm {erf}}( b^2(t)-b^2_{\rm max})] \, .
	\label{alpha_babcock}		  
   \end{eqnarray}
Apparently similar to system (1,2,3), 
this equation system has a number of
peculiarities: First, as suggested by \cite{Wilmotsmith2006},
there are two time delays in the system. 
The first one, $\tau_0$,  represents the time needed for meridional
circulation to transport the poloidal field, assumed to 
be produced by some Babcock-Leighton effect 
close to the solar surface, to the tachocline region, 
which is supposed to be the
site of the $\Omega$-effect. The second delay, 
$\tau_1$, represents the rise time
of flux-tubes which, in turn, transport
toroidal field from the tachocline region to 
the surface.
The effective $\alpha$-effect is supposed to work only between
a minimum $|b_{\rm min}|$ and a maximum value $|b_{\rm max}|$, where
flux-tubes start to rise when they have grown to 
a minimum strength $|b_{\rm min}|$, and $|b_{\rm max}|$
is the field amplitude at which the quenching of $\alpha$
becomes significant. 
Recently, this model has been extended by \cite{Hazra2014} who
added a stochastic fluctuation of $\alpha$ and argued that
the Babcock-Leighton mechanism alone cannot recover 
the solar cycle from a grand minimum.

What is new in our model, compared to that of \cite{Wilmotsmith2006}, 
is the consideration of either $b_{\rm min}$ or 
$\Omega$ as periodically time-dependent. Again, we have 
in mind a gravitational
perturbation with a period of 11.07 years (and some relative amplitude
${\mathcal A}$) as produced 
by the Venus-Earth-Jupiter system, i.e.
\begin{eqnarray}
b_{\rm min}(t)=b_{\rm min,0}(1+{\mathcal A} \sin{(2 \pi t/T_{\rm tidal})})
\end{eqnarray}
or
\begin{eqnarray}
\Omega(t)=\Omega_0 (1+{\mathcal A} \sin{(2 \pi t/T_{\rm tidal})}) \; .
\end{eqnarray}

Why consider $b_{\rm min}$ as time-dependent? The idea traces back
to \cite{Abreu2012} who had argued that the overshoot layer, 
which coincides
spatially with the tachocline, is crucial for the
storage and amplification of the magnetic flux tubes before they
eventually erupt to the solar photosphere.
The key factor here is the superadiabaticity, 
a dimensionless measure
of the stratification of the specific entropy. 
The maximum field strength of a flux tube that can be stored
prior to eruption is very
sensitive to this superadiabaticity; hence small changes of 
it (as provoked, e.g., by tidal forces) could 
decide about the ultimate strength 
of the rising flux tube. In our model this will
be emulated by a periodic time dependence of $b_{\rm min}$.

While such a periodic 
variation of the adiabaticity, and therefore
of $b_{\rm min}$, is still speculative, a corresponding 
11 years oscillation of
$\Omega$ has indeed been observed in form of 
``torsional oscillations'' \citep{Brown1989,Howe2009}.
Although these are commonly discussed in terms of a
large-scale feedback of Lorentz forces 
(Malkus-Proctor effect  \citep{Malkus1975})
or  $\Lambda$-quenching \citep{Kitchatinov1994}, 
they might also result directly from
planetary influences \citep{Zaqarashvili1997}.  

In the following we will assess if, and under which conditions, 
the two time periodic variations of $b_{\rm min}$ or $\Omega$ 
might lead to a synchronization of the dynamo. 
In all considered regimes we
will find dynamos with a positive product of $\alpha$ and
$\Omega$ to undergo pulsations, while
dynamos with negative product of $\alpha$ and
$\Omega$ usually oscillate.

\subsection{Flux-transport dominated regime}

We start with a regime in which the flux-transport 
is fast compared to dissipation effects, i.e. 
$\tau > \tau_0+\tau_1$.
Both positive and negative products of $\alpha$ and
$\Omega$ will be considered. 

\begin{figure}[!ht]
\includegraphics[width=120mm]{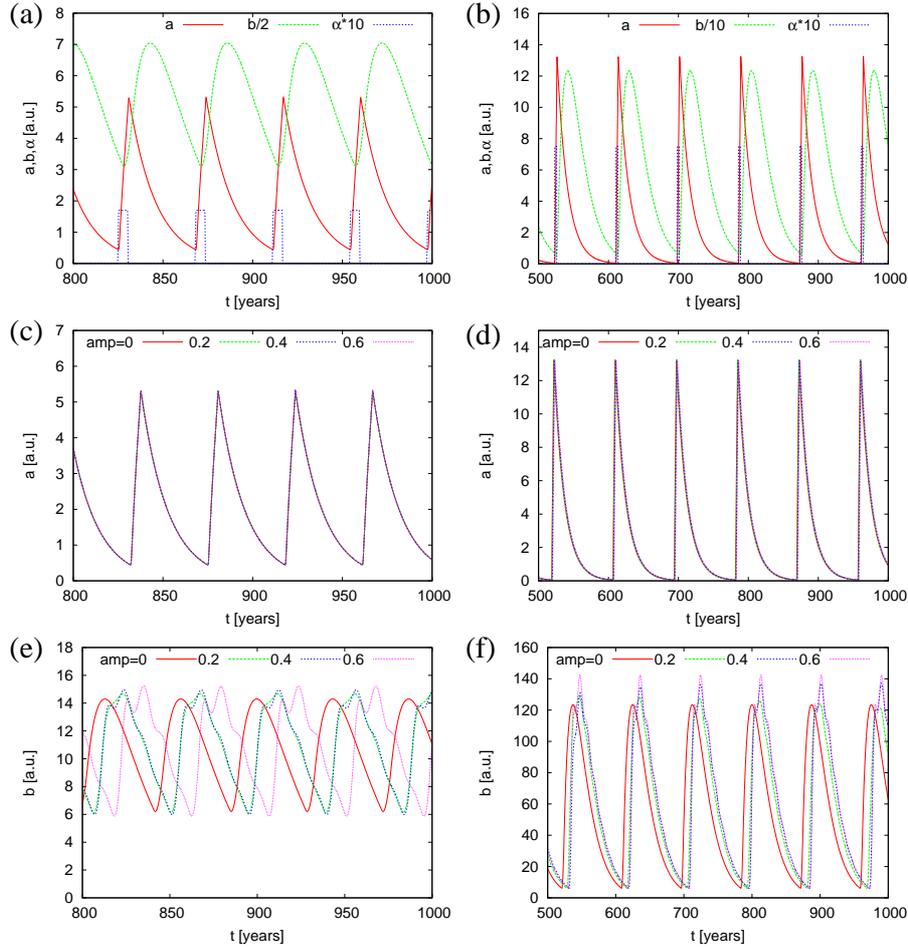}
\caption{Time series in the flux-transport dominated regime 
with positive product of $\alpha$ and $\Omega$.
The constant parameters are: 
$\tau=15$, $\tau_0=2$, $\tau_1=0.5$, $b_{\rm min}=1$, $b_{\rm max}=7$.
The variable parameters are:  $\alpha_0=0.17$, $\Omega=0.34$ (a,c,e),
$\alpha_0=0.75$, $\Omega=1.5$ (b,d,f).
Panels (a) and (b) show $a(t)$, $b(t)$ and $\alpha(t)$ for 
the unperturbed system (the behaviour is very 
similar to that shown in Fig. 7 of 
\cite{Wilmotsmith2006}).
Panels (c) and (d) show only $a(t)$ for 
various amplitudes $\mathcal{A}$ of the perturbation of 
$b_{\rm min}$ with perturbation period 11.07. Evidently, the 
periodic variation of $b_{\rm min}$ 
has a negligible effect. Panels (e) and (f)
show only $b(t)$ for various amplitudes $\mathcal{A}$ 
of the perturbation of 
$\Omega$. The variation of
the amplitude of $\Omega$ leads to some shifting effect, but 
no synchronization is observed.
}
\label{Fig:fluxtransport-positiv_alles}
\end{figure}

\subsubsection{Positive dynamo number}
Figure \ref{Fig:fluxtransport-positiv_alles} illustrates 
the results for two specific sets of dynamo parameters.
With the constant parameters 
$\tau=15$, $\tau_0=2$, $\tau_1=0.5$, $b_{\rm min}=1$,
$b_{\rm max}=7$, 
we evaluate  the two combinations
$\alpha_0=0.17$, $\Omega=0.34$ (a,c,e) and
$\alpha_0=0.75$, $\Omega=1.5$ (b,d,f).
Panels (a) and (b) show $a(t)$, $b(t)$ and $\alpha(t)$ for 
the unperturbed system which turns out 
very similar to the pulsation behaviour shown 
in Figure 7 of  \cite{Wilmotsmith2006}).
Panels (c) and (d) show (only) $a(t)$ for 
various amplitudes $\mathcal{A}$ of the 
perturbation of $b_{\rm min}$ with a period of 11.07. 
In either case, 
the effect of
periodic variation of $b_{\rm min}$ can be
neglected. Correspondingly, panels (e) and (f)
show (only) $b(t)$ for various amplitudes $\mathcal{A}$ 
of the perturbation of 
$\Omega$. Although some shifting effects are visible,  
there is no sign of any synchronization.
One might guess, however, that this has simply to do 
with the large gap between the period of the forcing 
(11.07) and the periods of the unperturbed dynamo, 
i.e., for ${\mathcal A}=0$,
which are very large (43 for (a,c,e) and 88 
for (b,d,f)).



\subsubsection{Negative dynamo number}

\begin{figure}[!ht]
\includegraphics[width=120mm]{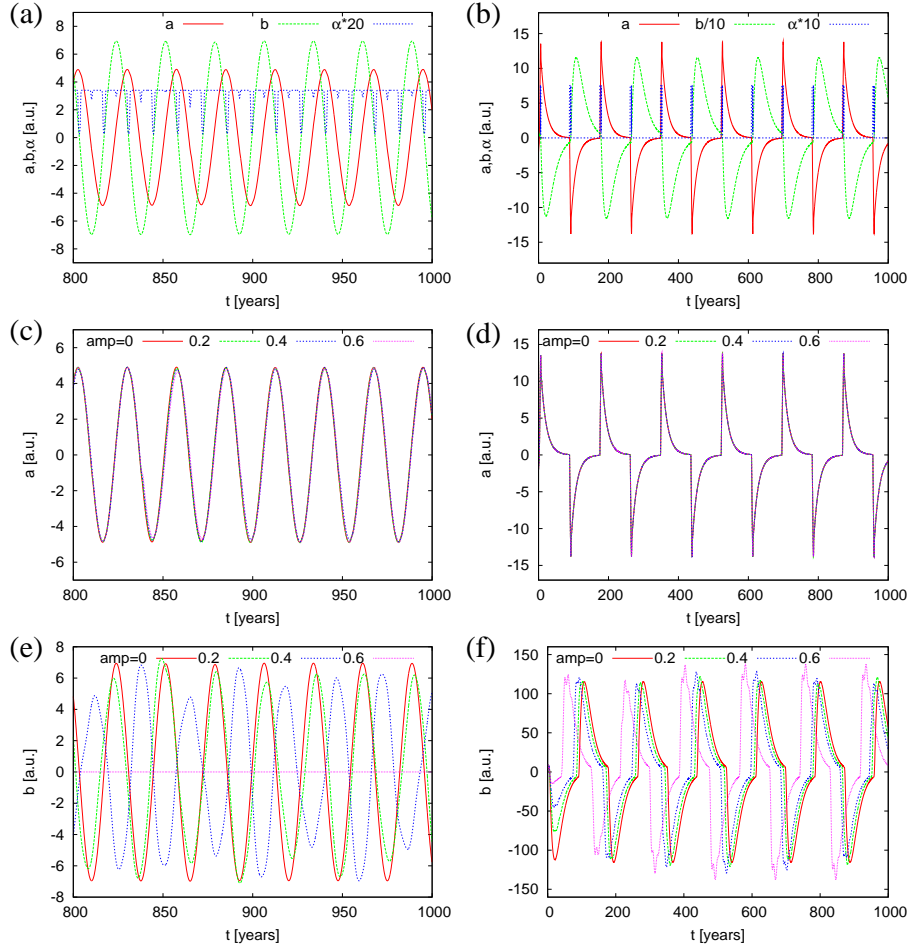}
\caption{Similar as Figure (\ref{Fig:fluxtransport-positiv_alles}) but
with negative product of $\alpha$ and $\Omega$.
The constant parameters are: 
$\tau=15$, $\tau_0=2$, $\tau_1=0.5$, $b_{\rm min}=1$, $b_{\rm max}=7$.
The variable parameters are:  $\alpha_0=0.17$, $\Omega=-0.34$ (a,c,e),
$\alpha_0=0.75$, $\Omega=-1.5$ (b,d,f).
Panel (a) corresponds to Figure 3, (b) to 
Figure 4 of \cite{Wilmotsmith2006}.
 Again, the 
periodic variation of $b_{\rm min}$ (c,d)
has a negligible effect.  The variation of
the amplitude of $\Omega$ (e,f) 
has some shifting effect, but 
there is no synchronization. Note the
switching off of the dynamo for $\mathcal{A}=0.6$ in (e).
}
\label{Fig:fluxtransport-negativ_alles}
\end{figure}

Keeping all parameters as before, and changing only the sign of $\Omega$,
we obtain the time series of Figure \ref{Fig:fluxtransport-negativ_alles}.
Rather than a pulsations we observe now the ``desired'' oscillation
of the dynamo.

Again, periodic variations of $b_{\rm min}$ have no 
noticeable effect (c,d). The periodic perturbation of 
$\Omega$ leads to some shifting, but not 
to any synchronization. This time, synchronization  
might have been expected for (c,e) since the unperturbed dynamo (a)
has a period of 29 which is not far from the double
of the excitation period. Interestingly, the dynamo
switches off completely for the largest perturbation 
$\mathcal{A}=0.6$ (e), where in its weak phase
$\Omega$ is reduced in such a way that the 
dynamo cannot work anymore.


\subsection{Diffusion dominated regime}

Now we consider the opposite case that the diffusion is 
faster than the flux transport, i.e $\tau < \tau_0+\tau_1$.

\subsubsection{Positive dynamo number}

\begin{figure}[!ht]
\includegraphics[width=120mm]{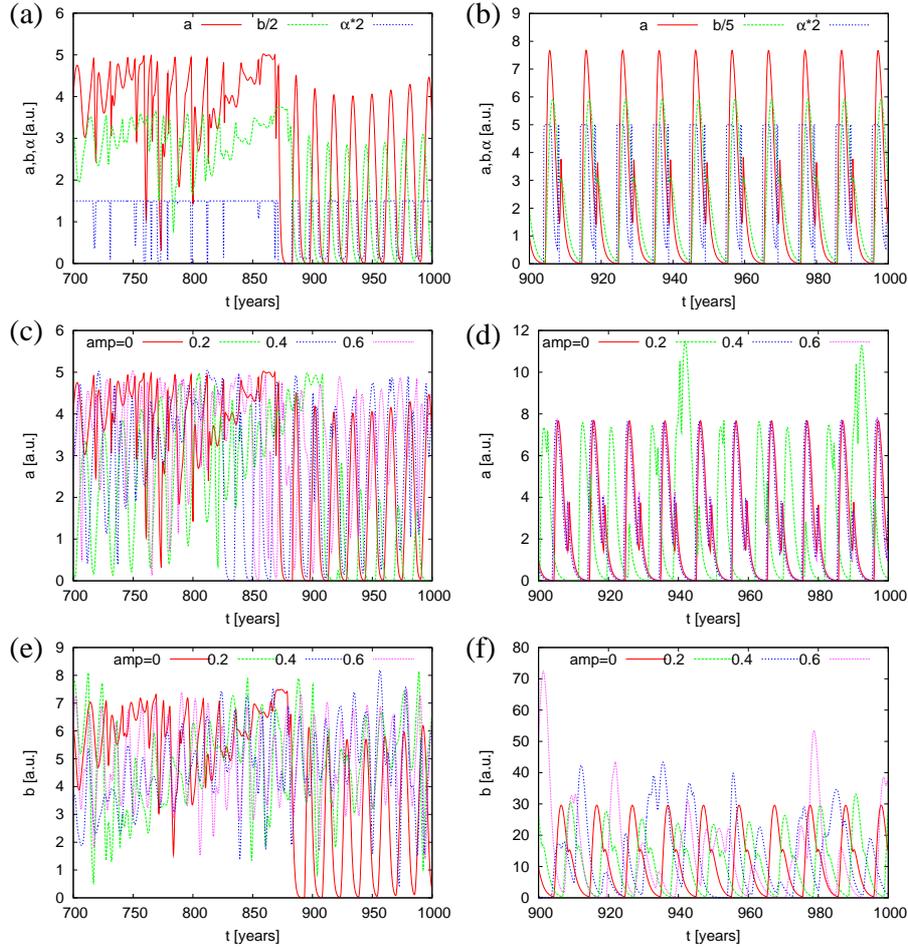}
\caption{Time series in the diffusion dominated regime 
with positive product of $\alpha$ and $\Omega$.
The constant parameters are:
$\tau=1$, $\tau_0=10$, $\tau_1=4$, $b_{\rm min}=1$, $b_{\rm max}=7$.
The variable parameters are:  $\alpha_0=0.75$, $\Omega=1.5$ (a,c,e),
$\alpha_0=2.5$, $\Omega=5.0$ (b,d,f).
Panels (a) and (b) show $a(t)$, $b(t)$ and $\alpha(t)$ for 
the unperturbed system (the erratic behaviour at the beginning of (a) 
is similar to that of Figure 12 
of \cite{Wilmotsmith2006} which works at $\alpha_0=-1$, $\Omega=-3$).
Panels (c) and (d) show $a(t)$ for 
various amplitudes $\mathcal{A}$ of the perturbation of 
$b_{\rm min}$ with perturbation period 11.07. 
Panels (e) and (f)
show $b(t)$ for various amplitudes $\mathcal{A}$ 
of the perturbation of 
$\Omega$.
The 
periodic variations of $b_{\rm min}$ and $\Omega$ 
provide some effects, but no synchronization.
}
\label{Fig:diffusion-positiv_alles}
\end{figure}

We start again with positive product of $\alpha$ and
$\Omega$.
Figure \ref{Fig:diffusion-positiv_alles} shows 
the results for two specific sets of dynamo parameters.
With the constant parameters 
$\tau=1$, $\tau_0=10$, $\tau_1=4$, $b_{\rm min}=1$, $b_{\rm max}=7$, 
we consider  the two combinations 
$\alpha_0=0.75$, $\Omega=1.5$ (a,c,e) and
$\alpha_0=2.5$, $\Omega=5$ (b,d,f).
Panels (a) and (b) show $a(t)$, $b(t)$ and $\alpha(t)$ for 
the unperturbed system. The erratic
behaviour at the beginning of (a) is actually similar 
to that of Figure 12 
of \cite{Wilmotsmith2006} (which has 
slightly different 
$\alpha_0=-1$, $\Omega=-3$, though).

For both dynamo strengths, 
neither the variation of $b_{\rm min}$ (c,d) nor that 
of $\Omega=5$ (e,f) lead to synchronization, despite some
shiftings and deformations of the signals.


\subsubsection{Negative dynamo number}

\begin{figure}[!ht]
\includegraphics[width=120mm]{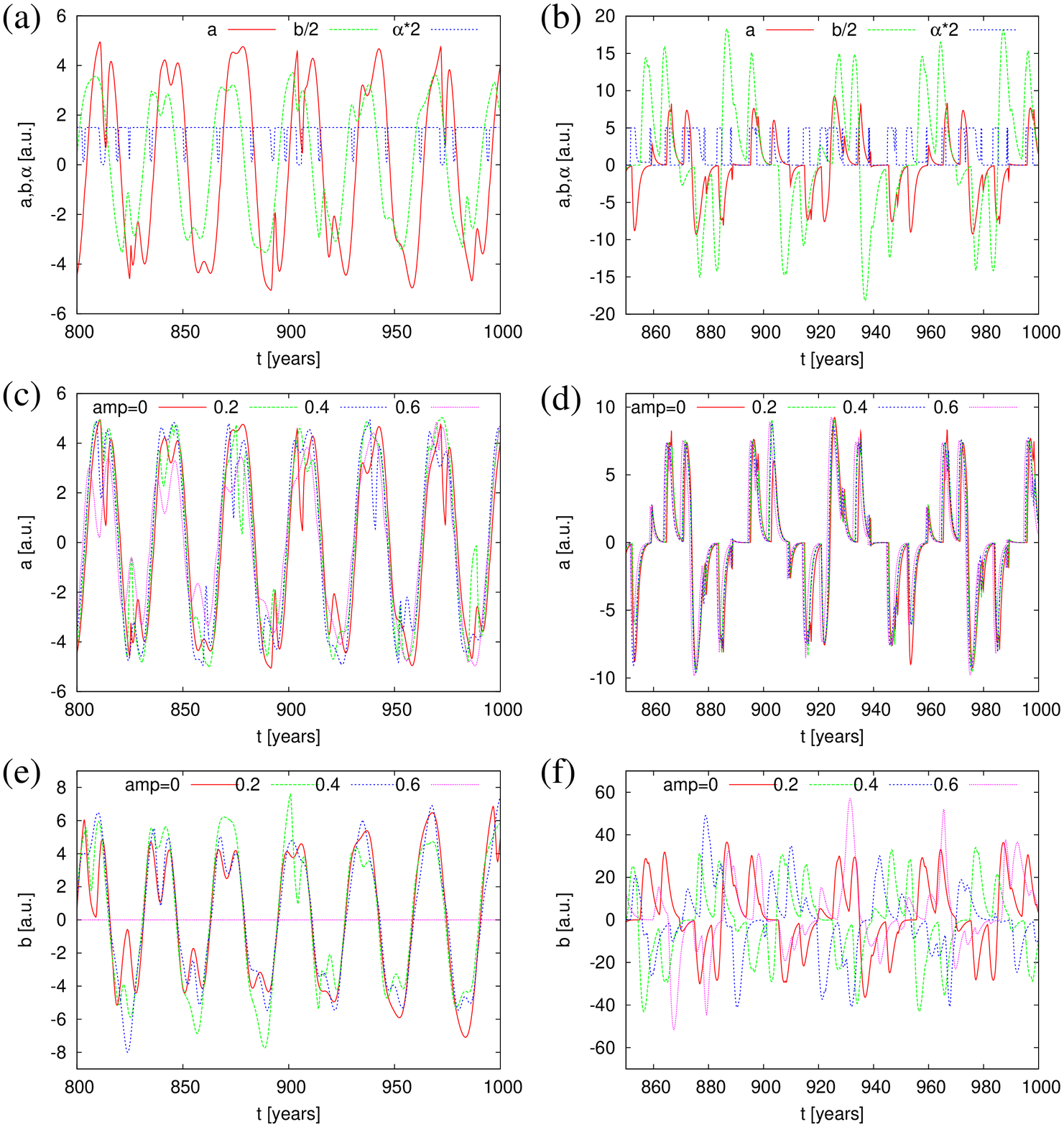}
\caption{Similar as Fig. (\ref{Fig:diffusion-positiv_alles}) but
with negative product of $\alpha$ and $\Omega$.
The constant parameters are:
$\tau=1$, $\tau_0=10$, $\tau_1=4$, $b_{\rm min}=1$, $b_{\rm max}=7$.
The variable parameters are:  $\alpha_0=0.75$, $\Omega=1.5$ (a,c,e),
$\alpha_0=2.5$, $\Omega=5.0$ (b,d,f).
The behaviour in (b) corresponds to that of Figure 8
of \cite{Wilmotsmith2006}.
The perturbation of 
$b_{\rm min}$ (c,d) and  $\Omega$ (e,f)
have some effect but do not lead to 
synchronization.
}
\label{Fig:diffusion-negativ_alles}
\end{figure}

We continue with the case of a negative product of $\alpha$ and
$\Omega$ in the diffusive regime. Again, Figure 
\ref{Fig:diffusion-negativ_alles} shows the typical oscillations
instead of the pulsations which had dominated for positive product.
With the constant parameters left unchanged,
the variable parameters are now:  $\alpha_0=0.75$, $\Omega=-1.5$ (a,c,e),
$\alpha_0=2.5$, $\Omega=-5.0$ (b,d,f).
The behaviour in (b) corresponds to that of Figure 8
of \cite{Wilmotsmith2006}.
The 
periodic variations of $b_{\rm min}$ (c,d) and $\Omega$ (e,f)
have some effect, but provide no real synchronization.

\subsection{Intermediate regime}

While the previous two regimes were already studied (without
the periodic perturbations) in \cite{Wilmotsmith2006}, we 
further consider here an intermediate regime characterized
by the time ordering $\tau_1 < \tau < \tau_0$, i.e. 
the diffusion is slower than the rise of flux-tubes, 
but faster than the meridional circulation.

\subsubsection{Positive dynamo number}

\begin{figure}[!ht]
\includegraphics[width=120mm]{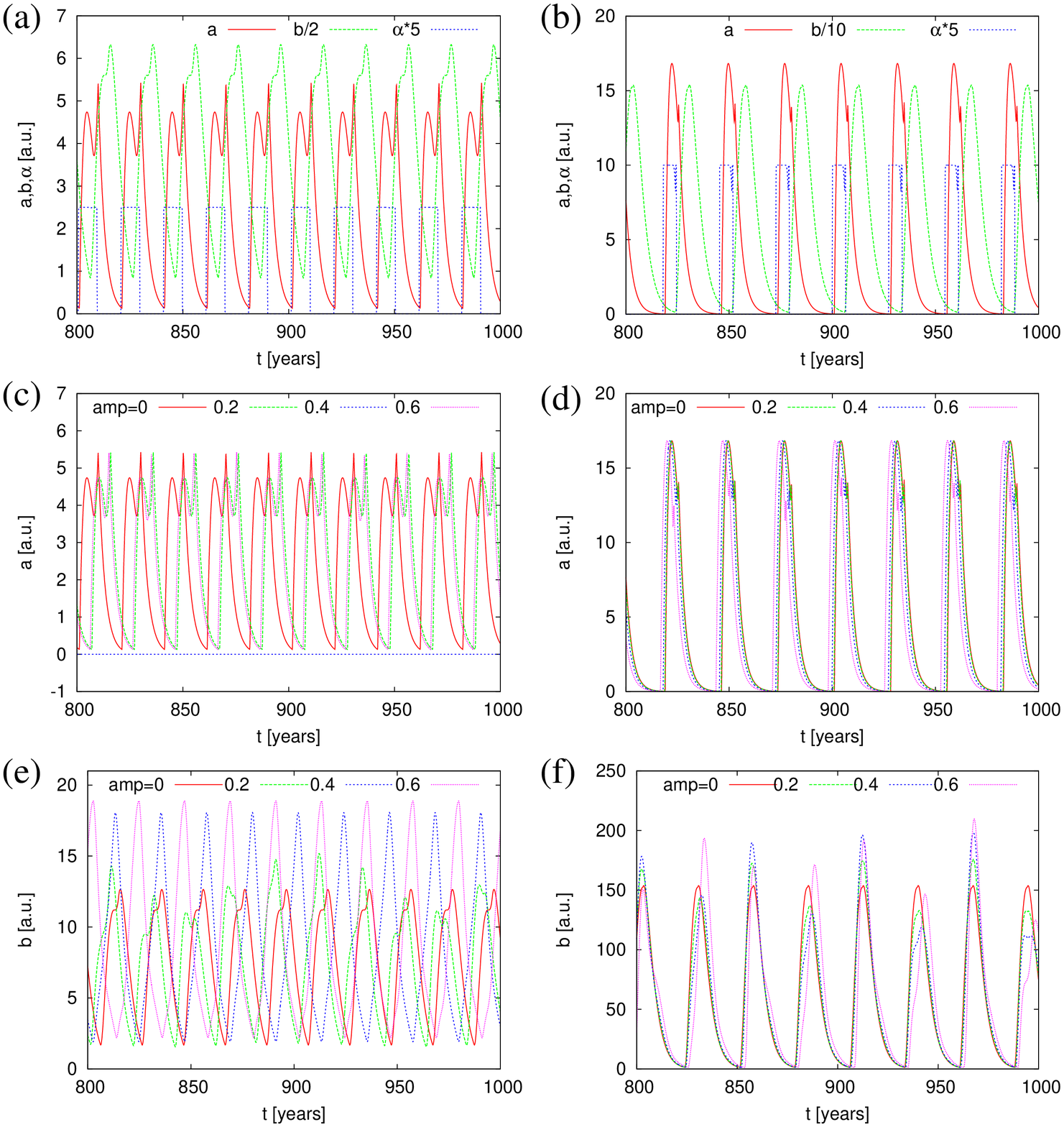}
\caption{Time series in the intermediate  regime 
with positive product of $\alpha$ and $\Omega$. 
The constant parameters are:
$\tau=3$, $\tau_0=5$, $\tau_1=1$, $b_{\rm min}=1$, $b_{\rm max}=7$.
The variable parameters are:  $\alpha_0=0.5$, $\Omega=1$ (a,c,e),
$\alpha_0=2$, $\Omega=4.0$ (b,d,f).
Panels (a) and (b) show $a(t)$, $b(t)$ and $\alpha(t)$ for 
the unperturbed system.
Panels (c) and (d) show (only) $a(t)$ for 
various amplitudes $\mathcal{A}$ of the perturbation of 
$b_{\rm min}$ with perturbation period 11.07. Panels (e) and (f)
show $b(t)$ for various amplitudes $\mathcal{A}$ 
of the perturbation of 
$\Omega$. The variation of
the amplitude of $\Omega$ leads to synchronization.
}
\label{Fig:intermediate-positiv_alles}
\end{figure}

Let us start again with positive product
of $\alpha$ and $\Omega$. With the constant 
parameters 
$\tau=3$, $\tau_0=5$, $\tau_1=1$, $b_{\rm min}=1$
$b_{\rm max}=7$, 
we check  the two combinations 
$\alpha_0=0.5$, $\Omega=1$ (a,c,e) and
$\alpha_0=2$, $\Omega=4$ (b,d,f).
Again panels (a) and (b) of Figure \ref{Fig:intermediate-positiv_alles}
show $a(t)$, $b(t)$ and $\alpha(t)$ for 
the unperturbed system which exhibit pulsations, as is 
usual 
for positive product of $\alpha$ and $\Omega$.
Panels (c) and (d) show  $a(t)$ for the perturbed system, with
various amplitudes $\mathcal{A}$ of the 
perturbation of $b_{\rm min}$. 
In either case (c) and (d), 
the effect of
periodic variation of $b_{\rm min}$ leads only
to a weak shift of the time series. 
Correspondingly, panels (e) and (f)
show $b(t)$ for various amplitudes $\mathcal{A}$ 
of the perturbation of $\Omega$. In (e) we observe 
for the first time a synchronization (compare the 
systematic 
shift of the maxima for $\mathcal{A}=0.6$ compared to those
for $\mathcal{A}=0$).

\begin{figure}[!ht]
\includegraphics[width=100mm]{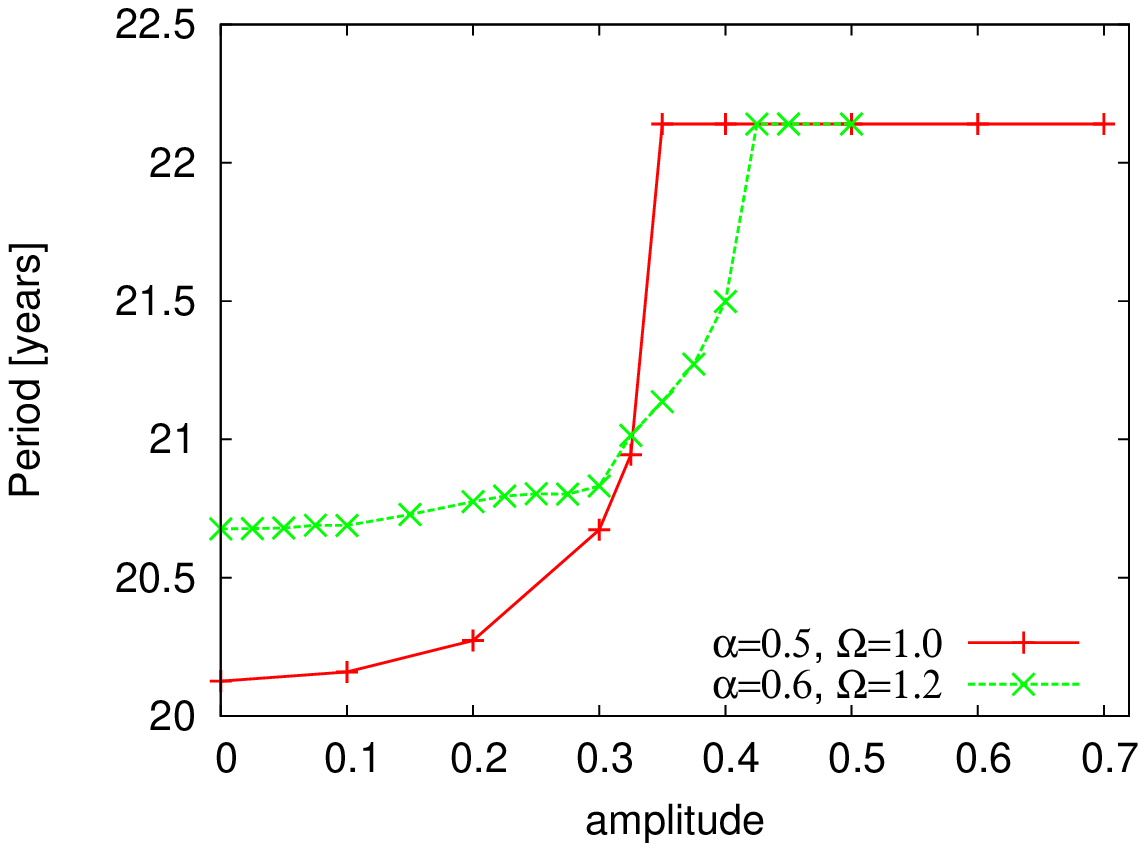}
\caption{Dominant period of pulsation in dependence
on the amplitude $\mathcal{A}$ 
of the $\Omega$ variation for
$\alpha_0=0.5$, $\Omega=1.0$ and $\alpha_0=0.6$, $\Omega=1.2$ in the
intermediate regime.
Synchronization
with twice the 11.07  driving period
appears at $\mathcal{A}=0.35$ and $\mathcal{A}=0.425$, respectively.
}
\label{Fig:intermediate-positiv-resonanzomega}
\end{figure}

This synchronization effect is further quantified 
in Figure \ref{Fig:intermediate-positiv-resonanzomega}.
For the parameters $\alpha_0=0.5$, $\Omega=1$, and
the close-by parameters $\alpha_0=0.6$, $\Omega=1.2$,
we demonstrate how the period of the dynamo
oscillation approaches the 
double period 22.14 of the 11.07 period variation of $\Omega$,
when the amplitude of the perturbation is increased.
What is observed here is, therefore, a synchronization 
of order 1:2 \citep{Pikovsky}. The typical
parabolic shape of the curves
is representative of the occurrence of parametric resonance
\citep{Giesecke2012,Giesecke2017}.
Still, the amplitude of the perturbation must be large
(around 0.4) in order to provide synchronization.
It is not very likely, that the typical 1 per cent changes 
of $\Omega$, as observed in the sun \cite{Howe2009},  
could provide such an entrainment effect.


\subsubsection{Negative dynamo number}

We switch over now to the case of negative product
of $\alpha$ and $\Omega$ und use, with all
other parameters unchanged,  
$\alpha_0=0.5$, $\Omega=-1$ (Figure \ref{Fig:intermediate-positiv_alles}(a,c,e)), 
and $\alpha_0=2$, $\Omega=-4$ (b,d,f).
We see that the stronger dynamo (b) undergoes a clear oscillation, 
whereas the weaker one (a) 
shows a behavior somewhere between oscillation and pulsation.
While the perturbations of the stronger dynamo (d,f) 
lead only to minor changes without synchronization, 
the perturbation of the weaker dynamo is more
interesting. From panel (e) we can try to
infer the length of the dominant 
period in more detail when varying $\mathcal{A}$
(Figure \ref{Fig:intermediate-negativ-resonanzomega}).
Although not as clear as in the previous Figure 
\ref{Fig:intermediate-negativ-resonanzomega}, we still 
obtain a sort of synchronization, this time already for
smaller values of $\mathcal{A} \sim 0.1$.

\begin{figure}[!ht]
\includegraphics[width=120mm]{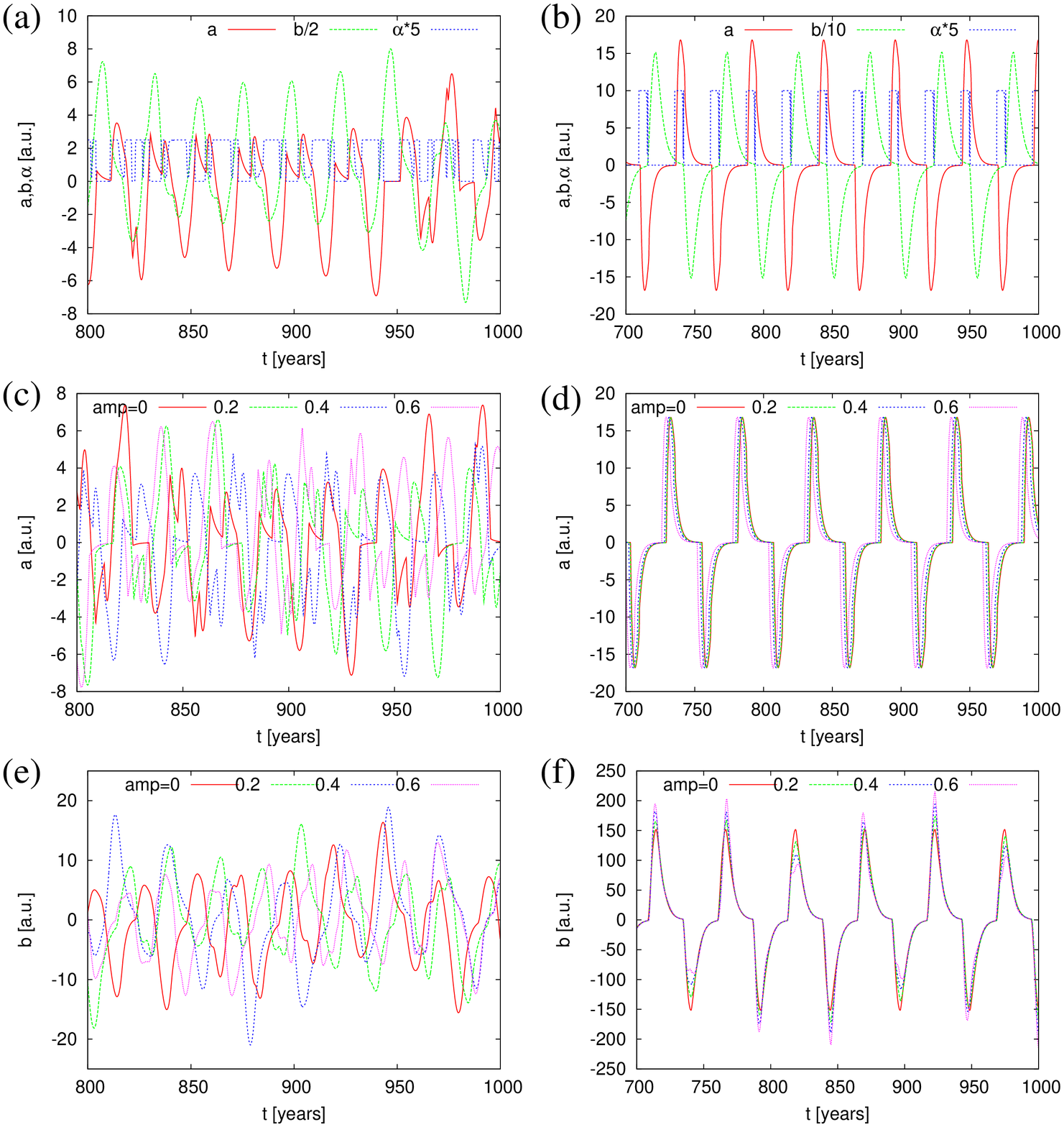}
\caption{Similar as Fig. (\ref{Fig:intermediate-positiv_alles}) but
with negative product of $\alpha$ and $\Omega$. The constant 
parameters are: $\tau=3$, $\tau_0=5$, $\tau_1=1$, $b_{\rm min}=1$, 
$b_{\rm max}=7$.
The variable parameters are:  $\alpha_0=0.5$, $\Omega=-1$ (a,c,e),
$\alpha_0=2$, $\Omega=-4.0$ (b,d,f).
This time, the variation of
the amplitude of $\Omega$ leads to synchronization for
the weak dynamo (e), while the strong dynamo (f) is by and large 
unaffected.
}
\label{Fig:intermediate-negativ_alles}
\end{figure}

\begin{figure}[!ht]
\includegraphics[width=100mm]{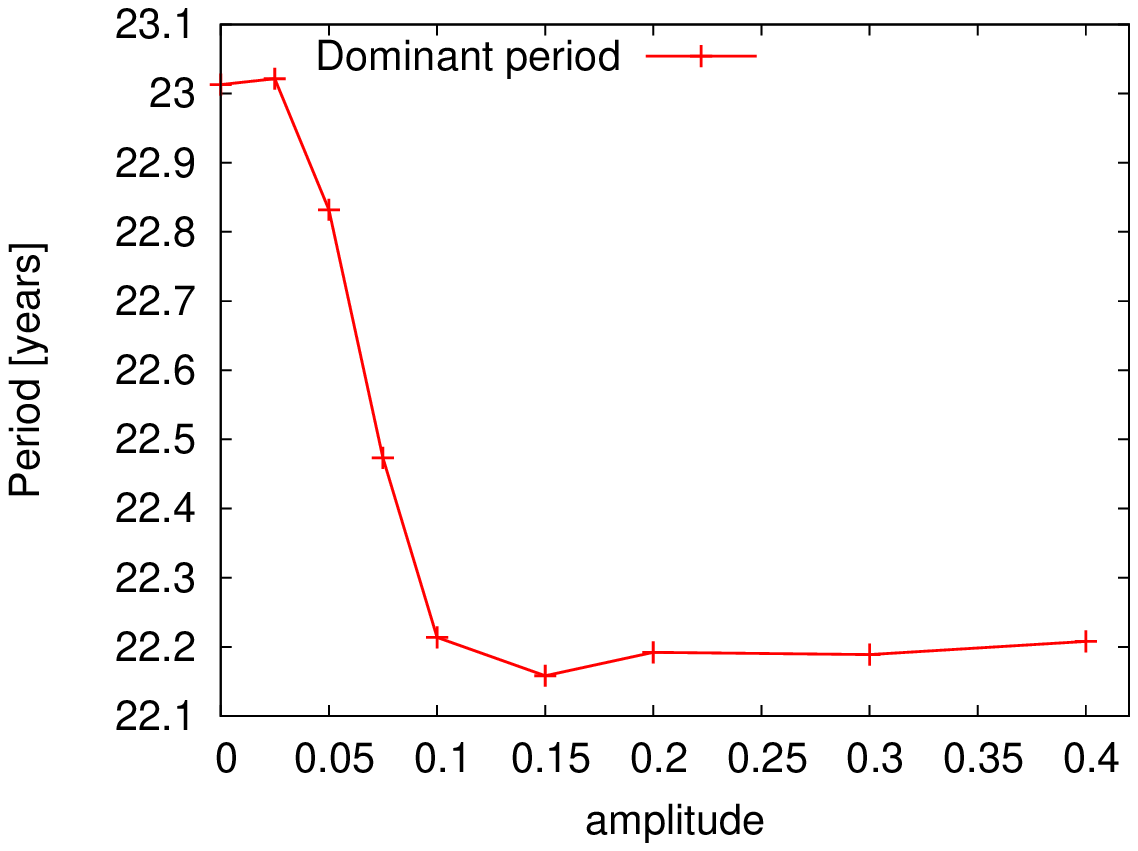}
\caption{Dominant period of oscillation in dependence
on the amplitude of the $\Omega$ variation for the 
intermediate regime with $\alpha_0=0.5$, $\Omega=-1$. 
Synchronization with twice the 11.07 years period appears 
at $\mathcal{A}=0.1$. Since the oscillation is slightly
irregular, the resonance is not as clearly expressed as in
Figure \ref{Fig:intermediate-positiv-resonanzomega}.}
\label{Fig:intermediate-negativ-resonanzomega}
\end{figure}

\section{Discussion and summary}

In his  statistical analysis,
titled "Is there a chronometer hidden deep 
in the sun?",
\cite{Dicke1978} had provided remarkable 
evidence for a positive answer to this 
fundamental question. 
 
Taking this finding seriously, 
we have assessed and compared 
the synchronizability of simplified
dynamo models of the Tayler-Spruit and the 
Babcock-Leighton type.
The synchronization of the Tayler-Spruit type dynamo
is based on the resonant excitation of the
$m=0$ component of the $\alpha$-effect (which
results from the  $m=1$ Tayler instability)
by an $m=2$ tidal-like perturbation.
We argued that a 
typical 11.07 years tidal perturbation,
as exerted by the conjunction cycle of
Venus, Earth and Jupiter,
 would end up 
in a 22.14 years 
oscillation of the dynamo field, at least 
for certain bands of the diffusion time $\tau$. For 
intervening bands, and for 
larger values of $\tau$, synchronization still occurs, 
but the 22.14 years oscillation is then replaced by
a 11.07 years pulsation. Whether these pulsations
are irrelevant for the sun, or whether they could be
linked to the behaviour during grand minima 
\citep{Weiss2016,Moss2017}, 
remains to be clarified in higher-dimensional models.
Remarkably, at any rate, is the phase coherence 
when passing from oscillations to pulsations, and back,
as evidenced in Figure \ref{Fig:maunder}.

In contrast to this, a corresponding Babcock-Leighton
type model proved rather  stubborn to
synchronization.
Specifically, we have pursued two ideas on 
how such a synchronization could work. The first 
one bears on the concept of a sensitive adiabaticity, i.e. 
flux storage capacity of the tachocline, which might be
easily influenced by  minor perturbations as 
provoked by tidal forces. This idea was 
implemented by periodically 
changing the value $b_{\rm min}$ which 
represents the critical threshold of the toroidal
field above which flux ropes would become magnetically 
buoyant.

The second idea took into account 
the - indeed observable - periodic change of the 
differential rotation, which is generally believed 
to be a {\it result} of the
dynamo-generated magnetic field (Malkus-Proctor effect
or $\Lambda$-quenching), 
but for which a direct planetary influence can not
be excluded completely.

Neither in the flux-transport dominated nor in the
diffusion-dominated regime we have observed any 
sign of synchronization. Only in the intermediate
regime, and here for comparably strong
perturbations of  $\Omega$, synchronization was 
obtained. 
For a positive product of $\alpha$ and
$\Omega$ it was possible to synchronize the
pulsations, for a negative product the
oscillation was synchronized. We stress, however,
that we have not covered  the relevant 
parameter space exhaustively.

Therefore, our results should in no way  be considered as an
argument against Babcock-Leighton dynamo models. 
They only underline the peculiarity 
of synchronizing
such types of dynamos by means of small periodic 
changes
of their governing physical parameters. This is 
in stark contrast to the amazingly simple and robust
synchronizability of
Tayler-Spruit type dynamos which need only 
a weak $m=2$ forcing to stoke the TI-related 
oscillations of $\alpha$. We recall here our 
previous
finding  \citep{Stefani2016} that the vacillations 
between left- and right-handed TI modes
do barely change the kinetic energy of the 
flow, so that not much energy is needed to 
excite them. The estimated tidal velocities of the 
order of 1\,m/s might serve this purpose well.
 
At any rate, it goes without saying 
that much more stringent
modeling is required to corroborate - or
reject - our TI-based synchronization model of the
solar dynamo.

\begin{acks}
This work was supported by 
Helmholtz-Gemeinschaft 
Deutscher For\-schungszentren (HGF)
in frame of the Helmholtz alliance LIMTECH.
\end{acks}

\section*{Disclosure of Potential Conflicts of Interest}
The authors declare that they have no conflicts of interest.

\end{article} 

\end{document}